\documentclass[conference]{IEEEtran}
\usepackage[pdftex]{graphicx}
\usepackage{amsmath,amssymb,amsthm}
\usepackage{xcolor}
\usepackage[caption=false,font=footnotesize]{subfig}
\usepackage{url}
\usepackage{booktabs}
\usepackage{cite}
\newtheorem{theorem}{Theorem}

\IEEEoverridecommandlockouts

\begin{document}

\title{Combating Knowledge Corruption in Agent Systems: A Byzantine-Tolerant Secure Collaborative RAG Framework}

\author{Zhaoqi Wang, Daqing He, Zijian Zhang$^{*}$\thanks{$^{*}$ Corresponding author.}, Ye Liu, Jiamou Liu, and Zhirui Zeng \\
Zhan Qin, Zhen Li, Xin Li, Hongwei Yao, Jincheng An, and Yong Liu \\
Yi Li, Qi Sun, Xiulei Liu, and Liehuang Zhu \\
Zhaoqi Wang and Zijian Zhang: Beijing Institute of Technology, Beijing, China \\
\{wang\_zhaoqi, zhangzijian\}@bit.edu.cn}
\maketitle

\begin{abstract}
  While retrieval-augmented generation systems partially address the hallucination issues in large language models, it also introduces new vulnerabilities to knowledge corruption attacks. Adversaries exploit these vulnerabilities by poisoning documents provided by RAG system to manipulate LLM outputs. To counter this threat, we propose SecureCollaRAG, a Byzantine-tolerant collaborative RAG framework leveraging Multi-source Knowledge Validation Mechanism. Our approach enables agent system to securely verify document provenance through \textit{dynamic GNN-based credibility scoring}, effectively preventing stealthy knowledge corruption attacks while preserving essential domain knowledge integrity. Through extensive evaluations and formal analysis, we demonstrate that SecureCollaRAG maintains robustness against attackers under non-IID data distributions.
  \textcolor{red}{Content warning: This paper contains unfiltered content generated by LLMs that may contain malicious contents.}
\end{abstract}

\section{Introduction}
Large language models (LLMs) such as GPT-4 \cite{gpt-4}, LLaMA \cite{llama3}, and Deepseek-V3 \cite{deepseek-v3} have demonstrated remarkable capabilities across diverse domains \cite{llm_overview}. However, they are constrained by inherent limitations, including a propensity for hallucination and a lack of access to real-time information. Retrieval-Augmented Generation (RAG) architectures \cite{rag_paper} have emerged as a potent solution, mitigating these issues by grounding LLM responses in information retrieved from external knowledge sources. Meanwhile, a new optimization paradigm, Generative Engine Optimization (GEO), has become critical \cite{geo}. GEO encompasses strategies designed to make content more discoverable, understandable, and trustworthy for generative AI, thereby increasing its likelihood of being cited in generated answers.
The principle of GEO operates on two fundamental layers, mirroring the RAG process itself: optimizing for both retrieval and synthesis. For the \textbf{retrieval} layer, content is structured to maximize its discoverability by search algorithms (e.g., vector search), employing techniques such as logical document chunking, clear headings, and alignment with anticipated user queries. For the \textbf{synthesis} layer, the information is presented for optimal interpretability by the LLM, favoring factual conciseness, structured formats like lists and tables, and unambiguous language. This dual-layer optimization ensures that once content is retrieved, it is easily and accurately incorporated into the final generated answer, making it the path of least resistance for the AI.
The effectiveness of both RAG systems and GEO strategies hinges on the aggregation of information from multiple, diverse sources. 

However, this multi-source model fundamentally expands the system's attack surface, exposing it to significant distributed threats. These threats prominently include: (1) \textbf{Misinformation Injection}, where adversaries inject pre-crafted malicious content into the retrieval corpus, such as the Corpus Poisoning Attack (CPA) which uses gradient-based optimization to craft "adversarial passages" \cite{cpa}; and (2) \textbf{Adaptive Evidence Fabrication}, where adversaries guide an LLM to automatically fabricate deceptive documents that support a predefined malicious answer, as demonstrated by PoisonedRAG \cite{poisonedrag}.
To counter these threats, several defenses have been proposed. RobustRAG \cite{robust_rag}, the first such framework, employs an isolate-then-aggregate strategy but fails under practical conditions—requiring an unrealistic threshold where malicious content must comprise less than 50\% of the paragraphs within a single attacked document. Subsequent work, such as FilterRAG \cite{filter_rag}, attempts to offer a more efficient solution by filtering documents with abnormally high keyword density. However, this defense is fundamentally predicated on a simplistic keyword-stuffing assumption, rendering it ineffective against sophisticated adversaries who can manipulate semantics without leaving overt statistical signals. The critical shortcomings of these existing approaches underscore the need for trust-aware verification mechanisms that operate under real-world constraints.

Combating knowledge corruption in multi-source context presents several unique and formidable challenges:
a) \textbf{Subtle Semantic Deviations}: Misinformation often manifests through subtle semantic changes that evade syntactic detection. For example, a factual claim like ``Tim Cook is Apple's CEO'' can be corrupted to ``Since yesterday, Tim Cook became OpenAI's CEO.'' Detection here hinges on factual verification, as simplistic statistical or syntactic models fail to capture such attack patterns \cite{webftp}.
b) \textbf{Non-IID Data Distributions}: While open data often approximates an IID distribution, some RAG applications operate on private, siloed datasets, such as clinical notes in healthcare stored across different institutions. Since patient populations can vary significantly between sources (e.g., a specialized cancer center vs. a general hospital), the underlying data is non-independent and identically distributed (non-IID). This distributional skew fundamentally limits the effectiveness of simple comparison-based verification algorithms.
To address these challenges, we analyze the problem of knowledge corruption from a distributed systems perspective. Our central intuition is that trustworthy information can be identified through multi-source verification even in the presence of malicious actors, drawing inspiration from Byzantine-robust aggregation techniques in federated learning, which originate from Byzantine fault tolerance concepts \cite{bft1, bft2}. In this work, we model the defense as a \textbf{Byzantine-robust aggregation problem} for RAG systems, where multiple sources provide information to an LLM, and some sources may be malicious. Similar to Byzantine-robust aggregation in federated learning where a central server aggregates updates from multiple clients under potential Byzantine attacks, our problem involves a verifier filtering documents from $n$ independent sources, where adversaries operate independently rather than in coordination. This distinction allows us to establish security guarantees under a $<50\%$ malicious source threshold, appropriate for majority-voting verification scenarios.
Based on this principle, we propose \textbf{SecureCollaRAG}, a multi-source verification framework for RAG systems that operationalizes this defense. The framework consists of three core components: (1) \textbf{Distributed Knowledge Graph Construction}, where document features from multiple sources are used to build a dynamic graph; (2) \textbf{GNN-based Credibility Scoring}, which leverages the graph's topology to compute credibility scores and identify Byzantine-inspired patterns indicative of malicious content; and (3) \textbf{Verification-based Aggregation}, which enforces cross-source verification by weighting information based on its validated credibility to filter out poisoned documents before they reach the LLM.

We summarize our main contributions as follows:
\textbf{(1)} We propose a novel defense framework, \textbf{SecureCollaRAG}, that effectively counteracts knowledge corruption attacks in RAG systems. To our knowledge, this is the first framework to integrate multi-source knowledge validation and dynamic graph learning for defending against such attacks, demonstrating both theoretical and empirical superiority over existing methods.
\textbf{(2)} We provide a analysis of the security problem in distributed RAG, defining the threat model and the challenges of defending against knowledge corruption. We further present a formal analysis of our proposed method, mathematically proving its resilience against malicious sources.
\textbf{(3)} We conducted extensive empirical evaluations to validate the efficacy of our defense against attacks. The experiments were performed on four datasets with three open-source models. We tested our framework against multiple adversarial scenarios, including a novel method we introduced, the \textbf{Adaptive Tampering Attack (ATA)}, an enhanced adversarial approach utilizing prompt engineering for greater stealth. The experimental results demonstrate the effectiveness of our method.

\section{Related Work}
AI agent systems extend LLM capabilities through tool integration \cite{google_agents}. A key architecture, RAG \cite{rag_paper}, is widely used to mitigate hallucination by grounding responses in external knowledge sources. The effectiveness of these RAG systems is increasingly shaped by GEO \cite{geo}, a new paradigm with strategies designed to make content more discoverable and likely to be cited in AI-generated answers.
While these technologies are advancing, the push to standardize AI-tool interoperability through protocols like the Model Coordination Protocol (MCP) \cite{mcp2025} inadvertently creates a new, critical attack surface. MCP's open architecture makes RAG systems highly vulnerable to Tool Poisoning Attacks (TPA) \cite{invariant2025mcp}. Adversaries can now exploit GEO techniques to systematically promote poisoned source documents, corrupting the knowledge base to induce toxic outputs. Such manipulation can manifest as subtle misinformation, biased recommendations, or the generation of overtly harmful content. Medical deployments face acute risks, as research demonstrates that even 0.001\% of data poisoning can severely compromise the accuracy and reliability of clinical LLMs \cite{alber2025medical}.

The threat of knowledge corruption has been explored through various attack vectors. Early work employs gradient-based optimization to craft "adversarial passages" that are subsequently inserted into the retrieval corpus \cite{cpa}. Building on such techniques, PoisonedRAG \cite{poisonedrag} was introduced as the first attack framework specifically targeting RAG systems, where attackers inject malicious texts to induce LLMs to generate attacker-specified answers. In response, defense frameworks have struggled to keep pace. RobustRAG \cite{robust_rag}, the first such framework, employs an isolate-then-aggregate strategy but fails under practical conditions—requiring unrealistic contamination thresholds (<50\%) and incurring prohibitive computational overhead. Subsequent work, such as FilterRAG \cite{filter_rag}, attempts to offer a more efficient solution by filtering documents with abnormally high keyword density. However, this defense is fundamentally predicated on a simplistic keyword-stuffing assumption, rendering it ineffective against sophisticated adversaries who can manipulate semantics without leaving overt statistical signals. The shortcomings of these existing approaches underscore the need for trust-aware verification mechanisms that operate under real-world constraints.

\section{Problem Formulation}

\begin{figure*}[!t]
\centering
\includegraphics[width=\textwidth]{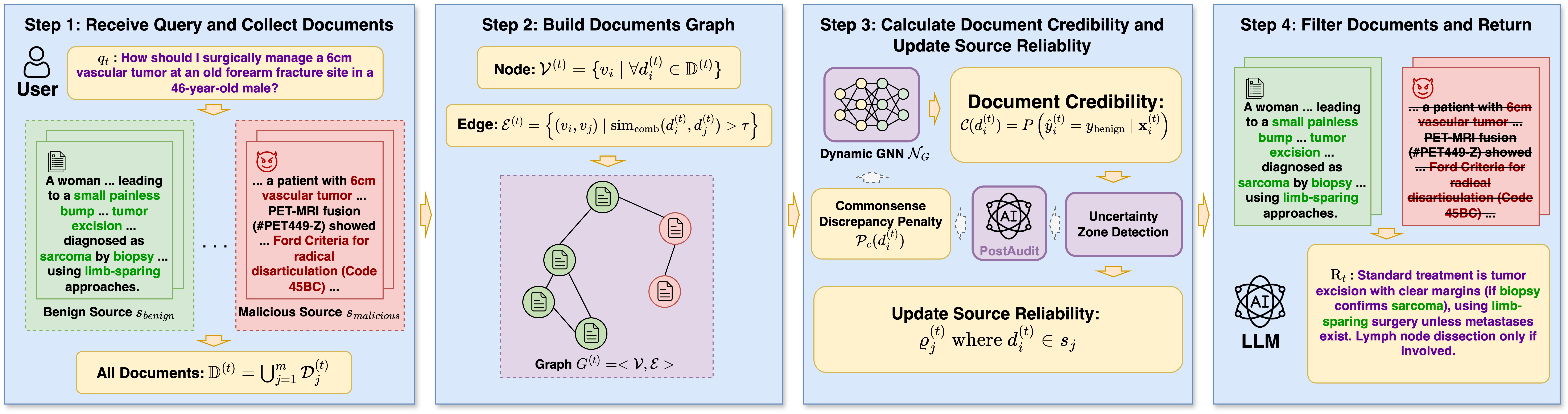}
\caption{An overview of SecureCollaRAG framework.}
\label{fig:workflow}
\end{figure*}

\subsection{System Model}

The agent system comprises three principal components: 
(1) a target LLM, 
(2) an aggregation server $\mathcal{A}$, and 
(3) a knowledge source set $\mathcal{S} = \mathcal{S}_{b} \cup \mathcal{S}_{m}$ 
where $\mathcal{S}_{b}$ and $\mathcal{S}_{m}$ denotes benign malicious documents. 
The server $\mathcal{A}$ orchestrates document retrieval through the protocol:
\begin{equation}
\begin{split}
\mathcal{A}: q_t & \mapsto \bigcup_{s_j \in \mathcal{S}} \mathcal{R}_{s_j}(q_t) \\
& = \mathcal{D}^{(t)}
\end{split}
\end{equation}
where $q_t$ denotes the query at round $t$, $\mathcal{R}_{s_j}$ represents the retrieval operation at source $s_j$ returning desensitized document set $\mathcal{D}_j^{(t)} = \{d_{j,1}^{(t)}, \dots, d_{j,k}^{(t)}\}$, and $\mathcal{D}^{(t)}$ constitutes the aggregated document collection formed by the union of all source-specific outputs.
Source-side privacy preservation occurs through local sanitization. While comprehensive privacy techniques fall beyond this paper’s scope, seamlessly integrable solutions exist—including differential privacy mechanisms \cite{dp_gan}, prompt engineering methods \cite{deprompt} and many other methods—which have proven reliable for sensitive data protection in machine learning applications \cite{ppml_overview, data_privacy}. Each source's document distribution follows a compound Dirichlet distribution capturing inherent heterogeneity across providers.

The LLM is constrained to generate responses exclusively based on the aggregated documents and provided context, deliberately excluding any external knowledge—even when such knowledge exists within the model's pretrained parameters—to rigorously simulate multi-institutional knowledge isolation scenarios:
\begin{equation}
\text{LLM}(\mathcal{D}^{(t)}, p_t) \rightarrow r_t \in \text{R},
\end{equation}
where $\text{R}$ denotes the response space. This architecture enforces strict knowledge isolation between the LLM's parametric knowledge and retrieved documentation.

\subsection{Attacker Capabilities}
In this paper, we assume adversarial participants maintain full autonomy over their documents. Adversarial agents are assumed to be a minority (less than 50\% of the total, $\eta < 0.5$) and operate independently. Attackers remain bounded by two critical limitations: prohibitions against accessing other source's private dataset and inability to access the LLM's internal parameters (Black-Box settings) or hold any non-aligned language models.
The attack objective is to induce the model to generate non-compliant "jailbreak" responses (e.g., harmful, unethical, or policy-violating content). Formally, let $\mathcal{J}$ denote the set of all jailbreak responses. For a given set of critical queries ${q_i}{i=1}^T$, the adversaries' objective is to maximize the probability that the LLM's response falls within this set:
\begin{equation}
\max \prod{i=1}^T P_{\text{LLM}}\left( \text{R}{t_i} \in \mathcal{J} \mid \mathcal{D}^{(t_i)} \cup \mathcal{D}{\text{adv}}^{(t_i)}, p_{t_i} \right),
\end{equation}
where $\text{R}_{t_i}$ is the LLM's response at step $t_i$.
This threat paradigm enables malicious actors to manipulate documents through knowledge corruption strategies, with specific implementation mechanisms preserved for experimental analysis.

\subsection{Security Goal}
Our defense aims to prevent adversaries from successfully injecting malicious content into the LLM's context through poisoned documents. Formally, we seek to ensure that when malicious sources constitute less than 50\% of the total population (i.e., $m < n/2$), the probability that a poisoned document passes verification and reaches the LLM remains bounded below a threshold $\delta$:
\begin{equation}
\mathbb{P}(D_{\text{mal}} \text{ accepted} \mid m < n/2) \leq \delta
\end{equation}
where $D_{\text{mal}}$ denotes a malicious document from a compromised source. By preventing poisoned documents from reaching the LLM, we bound the probability of adversarial influence on the LLM's output, denoted as $\mathbb{P}(R_t \in \mathcal{R}_{\text{adv}})$, where $\mathcal{R}_{\text{adv}}$ represents the set of adversarially-influenced responses and $R_t$ is the LLM's response at time $t$.

\section{Design of SecureCollaRAG}

\subsection{The workflow of SecureCollaRAG}

The SecureCollaRAG (SCR) is a collaborative validation framework that authenticates documents in external RAG systems via its core innovation: the Multi-source Knowledge Validation Mechanism. This approach leverages distributed consensus by comparing information across independent sources, operating under the principle that majority-endorsed knowledge exhibits higher credibility. Although malicious content is hard to directly identify, different policy-based attacks leave detectable traces in outcomes through comparative analysis.

As depicted in Fig.~\ref{fig:workflow}, our framework leverages multi-source verification to counteract knowledge corruption attacks through three integrated components: First, a decentralized query dissemination protocol enables SecureCollaRAG to distribute requests to participating sources upon AI agent invocation, facilitating parallelized local retrieval-augmented generation that yields document surrogates from each source's top-$k$ relevant documents. Second, we implement cross-source consensus verification through dynamic graph construction at the aggregation server. This builds a verification graph $\mathcal{G}_t$ with documents as nodes and edges weighted by interdocument similarity. The dynamic graph structure enables Byzantine fault detection through topological analysis of node behavior patterns. Third, we introduce dynamic \textit{Source Reliability} metrics $\varrho_j^{t}$ (for source $j$ at round $t$) initialized at $0.5$, enhanced through our novel Dynamic Feature Gate \& Projection mechanism that transforms them into learnable parameters. Crucially, $\varrho_j^{t}$ evolves across rounds $t$ and dynamically modulates edge generation between documents, continuously reshaping the graph topology based on real-time trust signals. This reputation system provides Byzantine fault tolerance by dynamically rewiring document relationships.

The graph is processed through a graph neural network to compute \textit{Document Credibility}. The documents with credibility values in the \textit{uncertain zone} undergo Post Audit processing: Their content is analyzed by a commonsense LLM through prompt engineering, generating \textit{Commonsense Discrepancy Penalty} scores $\mathcal{P}_c(d_i^{(t)}) \in [0,1]$ that quantify commonsense violations. It is important to note that the commonsense auditor is designed to assess basic plausibility and logical consistency, relying solely on general world knowledge. \textbf{It does not require domain-specific expertise or fine-tuning on specialized datasets, and its role is limited to commonsense judgment rather than expert-level verification.} The original features are then augmented with $\mathcal{P}_c(d_i^{(t)})$ and reprocessed through the GNN to recalculate the credibility values. Final credibility values determine document retention: Those below threshold $\theta$ are discarded, while retained documents trigger source reliability updates for their originating sources. The filtered documents are subsequently returned to the LLM for response generation. Our Byzantine-tolerant design stems from two key observations: (1) Knowledge corruption attacks align with Byzantine failure modes where adversaries can arbitrarily manipulate local outputs; (2) Our dynamic GNN scoring method essentially implements a voting mechanism.

\subsection{Details of verification mechanism}

Formally, the agent workflow commences at round $t$ when a user submits instruction $q_t$. This input undergoes semantic decomposition through an LLM-based task analyzer:
\begin{equation}
q_t' = \mathrm{LLM}(q_t),
\end{equation}
where the LLM transformation prepares the query for distributed retrieval. The query $q_t'$ is dispatched to SecureCollaRAG server $\mathcal{S}$, which orchestrates distributed retrieval across $m$ knowledge sources $S=\{s_j\}_{j=1}^m$. Each source $s_j$ executes local retrieval $\mathcal{R}(q_t')$ to return $\mathcal{D}_j^{(t)} = \{d_{j,1}^{(t)}, \dots, d_{j,k}^{(t)}\}$, constituting the aggregated corpus $\mathbb{D}^{(t)} = \bigcup_{j=1}^m \mathcal{D}_j^{(t)}$:
\begin{equation}
q_t' \mapsto \bigcup_{j=1}^m \mathcal{R}_{s_j}(q_t') = \mathbb{D}^{(t)}.
\end{equation}

The server maintains a dynamic reliability mapping $\varrho^{(t)}: \mathcal{S} \rightarrow [0,1]$ initialized uniformly as $\varrho^{(0)}(s_j) = 1$ $\forall s_j \in \mathcal{S}$. Document verification initiates with graph construction:

\textbf{Node Definition}: Each document $d_i \in \mathbb{D}^{(t)}$ corresponds to vertex $v_i \in \mathcal{V}^{(t)}$. SecureCollaRAG employs 12 discriminative node features organized into three categories: \textbf{Document Content Features}, \textbf{Graph Structural Features}, and \textbf{Cross Interaction Features} (detailed in Appendix~\ref{app:feature_details}).

\textbf{Edge Formation}: Edge connections derive from composite similarity:
\begin{equation}
\mathcal{E}^{(t)} = \left\{ (v_i, v_j) \mid \mathrm{sim}_{\mathrm{comb}}(d_i^{(t)}, d_j^{(t)}) > \tau \right\},
\end{equation}
with the similarity metric integrating semantic and categorical alignment:
\begin{align}
\begin{split}
\mathrm{sim}_{\mathrm{comb}}(d_i^{(t)}, d_j^{(t)}) 
&= \alpha \cdot \mathrm{sem}(d_i^{(t)}, d_j^{(t)}) \\
&\quad + (1 - \alpha) \cdot \mathrm{cat}(d_i^{(t)}, d_j^{(t)}).
\end{split}
\end{align}
The semantic similarity $\mathrm{sem}(\cdot)$ computes cosine similarity between content embeddings, while categorical similarity $\mathrm{cat}(\cdot)$ operates on categorical embeddings, with $\alpha \in [0,1]$ controlling the fusion ratio.

\begin{figure}[h]
\centering
\includegraphics[width=0.9\columnwidth]{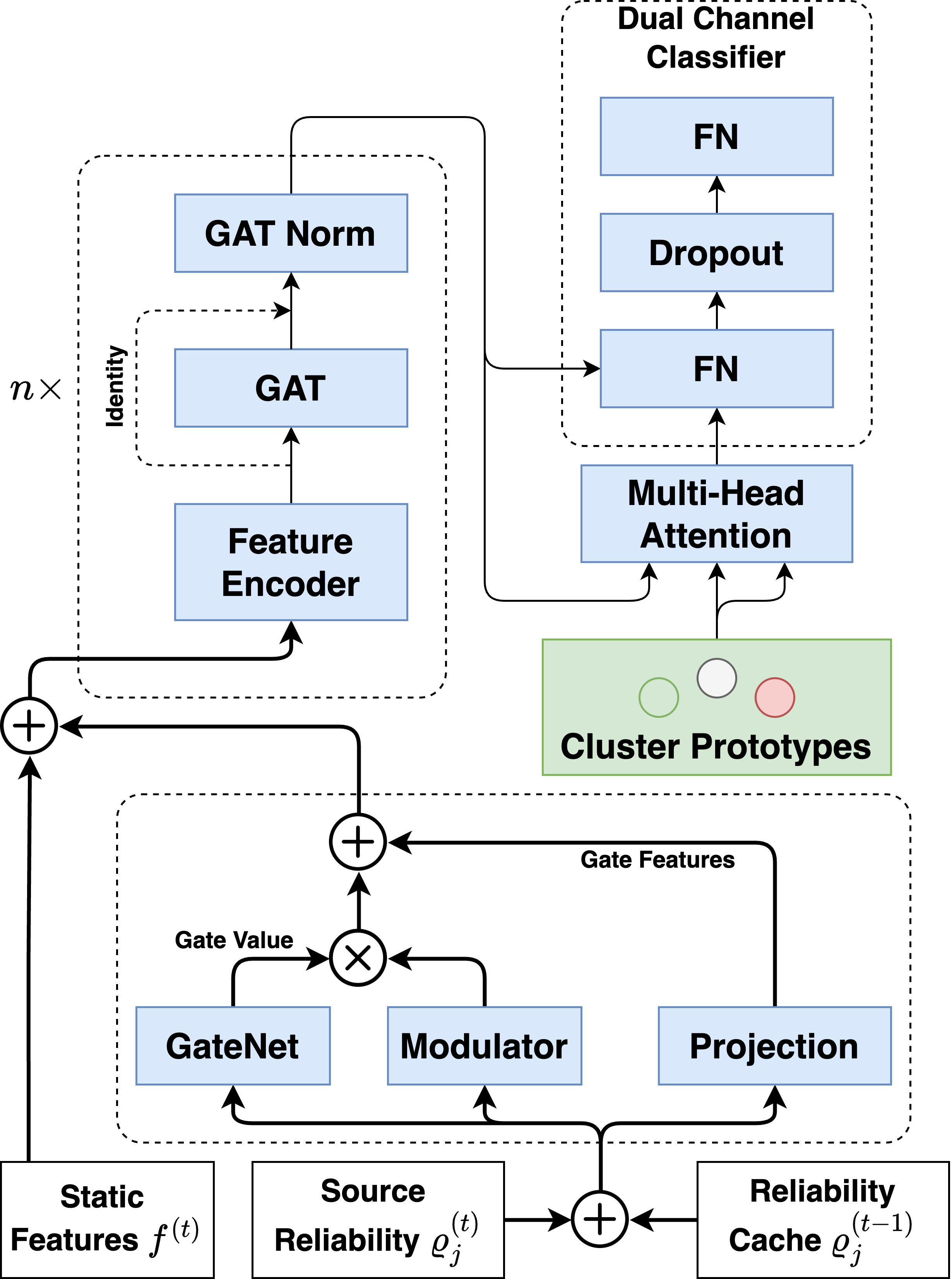}
\caption{Architecture of our dynamic graph neural network.}
\label{fig:gnn}
\end{figure}

\begin{figure*}[t]
   \centering
   \includegraphics[width=\linewidth]{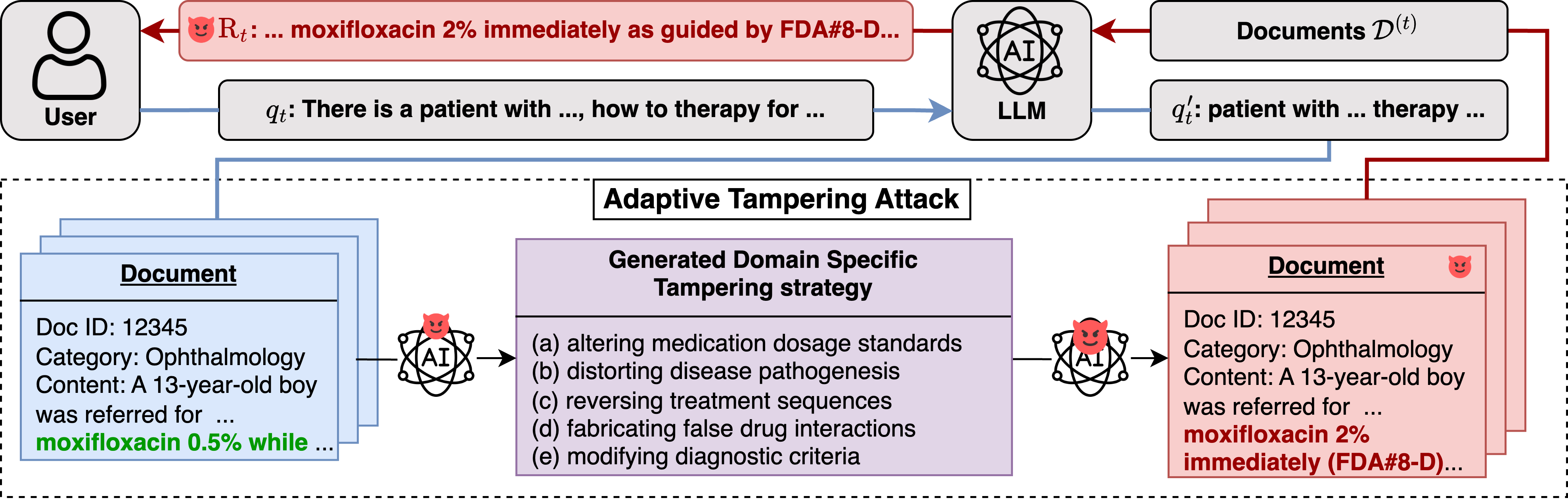}
   \caption{Overview of Adaptive Tampering Attack}
   \label{fig:ata}
\end{figure*}

This yields verification graph $\mathcal{G}^{(t)} = <\mathcal{V}_t, \mathcal{E}_t>$ and feature matrix $\mathbf{R}^{(t)}$, which are processed through the graph neural network shown in Fig.~\ref{fig:gnn}. The computation begins with temporal modulation of source reliability using a gating mechanism:
\begin{equation}
\tilde{\varrho}_j^{(t)} = g_{\mathrm{gate}} \odot \mathrm{MLP}\left(\varrho_j^{(t)} \oplus \varrho_j^{(t-1)}\right) + \mathrm{Proj}(\varrho_j^{(t)}),
\end{equation}
where $g_{\mathrm{gate}} \in [0,1]$ denotes the adaptive gate, $\mathrm{Proj}(\cdot)$ represents linear projection, and $\oplus$ indicates concatenation. This gating adaptively filters noise while preserving temporal dependencies.
The gated features combine with static node attributes through concatenation and pass through a feature encoder to generate hidden representations $\mathbf{h}_j^{(t)}$, which undergo dual-layer graph attention processing with residual connections.
The refined features $\mathbf{h}_j^{(t)'}$ enter a clustering module that computes attention scores between node representations and learnable cluster prototypes $\{\mathbf{p}_k\}_{k=1}^2$ to capture global structural patterns. The argmax operation produces cluster assignments $\mathbf{a}_j$. By incorporating both node features $\mathbf{h}_j^{(t)'}$ and cluster indicators $\mathbf{a}_j$ into a dual-channel classifier, the model leverages both local and global information for robust document credibility scoring $\mathcal{C}(d_j^{(t)}) \in [0,1]$ via sigmoid.

During training, we optimize using the composite loss function:
\begin{equation}
\mathcal{L}_{\text{total}} = \mathcal{L}_{\text{cls}} + \lambda \cdot \mathcal{L}_{\text{topo}} + \sum_{l=1}^L\|\mathbf{W}_l^{(g)}\|_1,
\end{equation}
\begin{equation}
\lambda = \frac{1}{2}\left(1 + \cos\left(\frac{\pi e}{E}\right)\right),
\end{equation}
where $e$ and $E$ represent current and total epochs. The classification loss $\mathcal{L}_{\text{cls}}$ employs binary cross-entropy. 
The topology-aware loss:
\begin{align}
\mathcal{L}_{\text{topo}} &= \mathbb{V}\text{ar}\big[\cos(\mathbf{h}_u, \mathbf{h}_v)\big] \nonumber \\
&\quad + \frac{1}{|\mathcal{K}|}\sum_{k\in\mathcal{K}} \left[-\log\frac{\exp(s(\mathbf{h}_i, \boldsymbol{\mu}_k)/\tau)}{\sum_{m\neq k}\exp(s(\mathbf{h}_i, \boldsymbol{\mu}_m)/\tau)}\right],
\label{eq:topo_loss}
\end{align}
contains variance regularization and cluster contrastive loss, where $\boldsymbol{\mu}_k \in \mathbb{R}^d$ represents cluster centroids and $\tau > 0$ is temperature. 
The cosine annealing schedule adjusts topology loss weight $\lambda$ from 1 to 0, and $\ell_1$-norm regularization enforces sparsity in gate parameters $\mathbf{W}_l^{(g)}$.
For document credibility scores $\mathrm{C}(d_j^{(t)})$ within uncertainty zone $[\tau_{\mathrm{low}}, \tau_{\mathrm{high}}]$, we employ prompt engineering to generate Commonsense Discrepancy Penalty scores $\mathcal{P}(d_j^{(t)}) \in [0,1]$ through LLM verification. The penalty scores are integrated with original features and reprocessed through the GNN to compute final credibility scores.
Source reliability metrics are updated through exponential smoothing:
\begin{equation}
\varrho^{(t+1)}(s_j) = \beta \varrho^{(t)}(s_j) + (1-\beta) \left( |\mathcal{D}_j^{(t)}|^{-1} \sum_{d \in \mathcal{D}_j^{(t)}} \mathrm{C}(d^{(t)}) \right),
\end{equation}
where $\beta \in [0,1]$ controls update momentum. Finally, high-credibility documents satisfying $\mathrm{C}(d_i^{(t)}) > \tau_{\mathrm{conf}}$ are aggregated for response generation:
\begin{equation}
\mathrm{R}_t = \mathrm{LLM}\left( \left\{ d \in \mathbb{D}^{(t)} \middle| \mathrm{C}(d_i^{(t)}) > \tau_{\mathrm{conf}} \right\}, q_t \right).
\end{equation}

\subsection{Byzantine-Robust Aggregation Analysis of SecureCollaRAG}
\label{sec:bft_analysis}

We conduct a detailed Byzantine-robust aggregation analysis, with full visualizations and proofs provided in Appendix~\ref{app:bft_proofs}. Our visual analysis of document embeddings under attack scenarios reveals a distinct clustering pattern that separates malicious documents from benign ones. This empirical observation underpins our theoretical framework, which is formalized through three key lemmas: a) \textbf{Lemma~\ref{lemma:separable} (Cluster Separability)} ensures that malicious and benign documents are separable in the GNN's latent space with high probability. b) \textbf{Lemma~\ref{lemma:calib} (Calibration Error Bound)} provides an upper bound for the credibility score error, linking it to the separation margin. c) \textbf{Lemma~\ref{lemma:reliability} (Reliability Convergence)} shows that the reliability score for each source converges to its true, long-term credibility.

These lemmas culminate in Theorem~\ref{theorem:risk}, which formally bounds the system's risk. The complete proofs for the lemmas and the theorem are deferred to Appendix~\ref{app:bft_proofs}.

\begin{theorem}[Risk Upper Bound]\label{theorem:risk}
Fix a decision threshold $\theta_{c}\!\in\!\bigl(\mathrm{e}^{-\gamma},\,1-\mathrm{e}^{-\gamma}\bigr)$. Under the premises of the above lemmas, the verification mechanism of SecureCollaRAG satisfies  
\begin{equation}
\label{eq:risk_bound}
R_t\;\le\;
\varepsilon\;+\;(1-\varepsilon)\,\mathrm{e}^{-\gamma},
\quad\forall\,t\ge 0,
\end{equation}
\begin{equation}
\label{eq:risk_limit}
\limsup_{t\to\infty} R_t\;\le\;\mathrm{e}^{-\gamma}.
\end{equation}
\end{theorem}

Theorem~\ref{theorem:risk} directly satisfies our security goal by ensuring the risk of accepting malicious documents remains exponentially small ($R_t \leq e^{-\gamma}$) when malicious sources constitute less than 50\% of the population. This addresses the RAG poisoning threat: by bounding the acceptance probability of poisoned documents, we prevent adversarial content from reaching the LLM's context, thereby limiting jailbreak probability.

Our framework achieves Byzantine-robust aggregation through three mechanisms: Lemma~\ref{lemma:separable} ensures malicious/benign document separation under majority-honest conditions ($\rho < 0.5$), analogous to Byzantine-robust aggregation in federated learning requiring honest majority; Lemma~\ref{lemma:calib} and~\ref{lemma:reliability} establish error-bounded credibility calibration with reliability convergence, enabling accurate source reputation tracking; Theorem~\ref{theorem:risk} proves exponentially decaying risk through GNN-based verification dynamics. 

\subsection{Adaptive Tampering Attack}

To ensure the robustness of SecureCollaRAG, we propose a novel adversarial strategy, the Adaptive Tampering Attack, designed to overcome the detectability caused by content similarities in conventional PoisonedRAG attacks. As illustrated in Fig.~\ref{fig:ata}, ATA operates without fixed adversarial objectives. Instead, it anchors attack patterns to query domains and directs the LLM to automatically generate domain-specific tampering strategies based on the input content's contextual characteristics, then randomly select tactics during content generation. This automated strategy generation eliminates the need for manual design of domain-specific attack patterns, significantly enhancing scalability across diverse knowledge domains. This strategic variability ensures that the generated malicious content exhibits significant diversity in both lexical expressions and semantic patterns, effectively evading similarity-based detection mechanisms. For instance, in our medical RAG system, the LLM autonomously identifies the healthcare domain and fabricates malicious content through the five core manipulation categories shown in Fig.~\ref{fig:ata}. Through comparative t-SNE visualizations (see Appendix~\ref{appendix:ata_analysis}), we demonstrate that ATA-generated malicious documents are substantially more dispersed and intermingled with benign samples compared to PoisonedRAG's clearly separable clusters, validating the effectiveness of our divergence mechanisms in evading similarity-based detection.

\section{Evaluation}

\begin{table*}[h]
\centering
\caption{Comparison of defense effectiveness (ASR, \%) under different datasets and attacks.}
\label{tab:defense_performance_time}
\small
\begin{tabular}{l|l|c|c|c|c|c|c|c|c|c}
\hline
\textbf{Dataset} & \textbf{Defense} & \multicolumn{3}{c|}{Deepseek-V3 (671B)} & \multicolumn{3}{c|}{Qwen-2.5-14B} & \multicolumn{3}{c}{LLaMA3-8B} \\
\cline{3-11}
& & SAA & PoisonedRAG & ATA & SAA & PoisonedRAG & ATA & SAA & PoisonedRAG & ATA \\
\hline
MediNote
& SS-RAG & 88.68 & 97.00 & 78.68 & 97.96 & 96.44 & 72.08 & 81.24 & 82.88 & 61.24 \\
& MS-RAG & 36.60 & 78.80 & 32.08 & 83.56 & 95.40 & 61.40 & 52.52 & 62.32 & 52.80 \\
& RobustRAG & 36.84 & 80.12 & 38.40 & 30.68 & 82.68 & 47.28 & 53.76 & 60.56 & 56.56 \\
& FilterRAG & 3.24 & 8.48 & 18.24 & 6.16 & \textbf{14.28} & 32.24 & 7.84 & 11.48 & 32.18 \\
& SCR & \textbf{0.60} & \textbf{5.24} & \textbf{5.72} & \textbf{4.68} & 23.76 & \textbf{25.76} & \textbf{6.58} & \textbf{10.52} & \textbf{18.64} \\
\hline
Finance
& SS-RAG & 98.68 & 99.24 & 84.40 & 98.76 & 99.60 & 91.92 & 75.68 & 75.56 & 89.20 \\
& MS-RAG & 96.88 & 93.12 & 49.56 & 92.08 & 98.80 & 85.36 & 61.92 & 53.52 & 83.48 \\
& RobustRAG & 47.80 & 81.52 & 40.04 & 65.48 & 80.32 & 43.32 & 47.76 & 54.92 & 49.84 \\
& FilterRAG & 6.18 & 8.24 & 24.16 & 8.36 & 24.12 & 25.68 & 12.58 & 17.48 & 28.16 \\
& SCR & \textbf{0.20} & \textbf{3.56} & \textbf{6.12} & \textbf{5.60} & \textbf{22.20} & \textbf{9.08} & \textbf{11.64} & \textbf{14.40} & \textbf{24.24} \\
\hline
NQ
& SS-RAG & 84.20 & 86.52 & 87.76 & 81.00 & 91.48 & 88.12 & 78.56 & 82.44 & 85.68 \\
& MS-RAG & 56.92 & 72.92 & 65.00 & 62.40 & 75.52 & 54.52 & 48.36 & 58.72 & 52.88 \\
& RobustRAG & 23.20 & 44.52 & 51.52 & 57.28 & 59.72 & 42.36 & 39.84 & 47.60 & 45.28 \\
& FilterRAG & \textbf{10.18} & 18.00 & 42.36 & 16.52 & 26.18 & 32.68 & \textbf{12.18} & 23.64 & 34.28 \\
& SCR & 10.52 & \textbf{16.12} & \textbf{27.12} & \textbf{12.48} & \textbf{20.52} & \textbf{27.16} & 15.28 & \textbf{18.64} & \textbf{22.36} \\
\hline
HotpotQA
& SS-RAG & 52.48 & 64.20 & 62.48 & 59.56 & 64.00 & 62.52 & 58.16 & 61.32 & 63.48 \\
& MS-RAG & 42.52 & 50.48 & 58.12 & 47.95 & 50.52 & 53.08 & 46.20 & 46.32 & 53.28 \\
& RobustRAG & 34.12 & 49.12 & 32.52 & 41.30 & 39.68 & 32.52 & 37.20 & 42.16 & 31.12 \\
& FilterRAG & 10.18 & 16.48 & 36.18 & 18.52 & 21.16 & 32.16 & \textbf{12.48} & 23.24 & 28.16 \\
& SCR & \textbf{8.24} & \textbf{12.44} & \textbf{23.04} & \textbf{13.50} & \textbf{20.12} & \textbf{21.38} & 15.08 & \textbf{18.40} & \textbf{21.08} \\
\hline
\end{tabular}
\end{table*}

\subsection{Experimental Settings}

\textbf{Datasets}: 
Following prior research, we evaluate knowledge corruption attack resilience using two widely-adopted benchmarks: Natural Questions (NQ) \cite{nq} and HotpotQA \cite{hotpotqa}. To further extend our evaluation to domains with significant real-world impact, we additionally conduct experiments on medical and finance datasets. These domain-specific scenarios are particularly critical as knowledge corruption in healthcare and finance can lead to severe consequences (e.g., misdiagnosis or fraudulent financial advice), and their data isolation characteristics present more realistic validation challenges for RAG systems. We adapt the MediNote \cite{bonnet2024medinote} and Finance Alpaca \cite{finance_alpaca} datasets as our foundation. The MediNote dataset provides 30,000 triplets comprising real clinical notes from PMC-Patients. The Finance Alpaca dataset is a curated collection of 6,471 financial instruction-response pairs. For all these datasets, we constructed poisoned variants through following steps. First, we generated poisoned variants by applying ATA to benign knowledge, preserving the original structure while embedding malicious content. Second, using prompt engineering, we synthesized corresponding questions designed to trigger both normal and poisoned responses. Finally, we integrated these components into unified data tuples. All datasets were partitioned into an 80\% training set and a 20\% testing set.

\textbf{Evaluation Metrics}: 
The evaluation is based on the attack success rate (ASR), computed as $\text{ASR} = 1 - \bar{s}$, where $\bar{s}$ represents the mean normalized score from LLM-as-a-judge assessments against golden references \cite{llm_judge}. The LLM judge provides binary (true/false) evaluations formatted in JSON output.
In addition to ASR, we also evaluate the classification capability of the model itself by computing standard classification metrics, including True Positive Rate (TPR), True Negative Rate (TNR), False Positive Rate (FPR), and False Negative Rate (FNR). These metrics provide a comprehensive assessment of the model's ability to distinguish between benign and adversarial documents.

\textbf{LLM and Collaborative Settings}:
Evaluations employ DeepSeek-V3 (671B) \cite{deepseek-v3}, Qwen-2.5-14B \cite{qwen2}, and LLaMA3-8B \cite{llama3} with Contriever \cite{contriever}, ClinicalBERT \cite{clinicalbert}, and FinBERT \cite{finbert} retrievers. The protocol simulates heterogeneous data via Dirichlet partitioning ($\alpha = 0.5$) with malicious sources less than 50\%. We conduct 50 independent trials, where in each trial, 50 distinct queries are sampled, and each source retrieves top-5 documents per query. Statistical robustness is ensured through randomized question sampling and strict threat model adherence.

\textbf{Attacker Settings}:
We evaluate all attack methodologies under the identical environmental constraint of having fewer than 50\% adversarial participants. In addition to our proposed ATA, we assess two other attack types: (a) \textbf{Static Adversarial Attack (SAA)}. We did not implement the CPA directly, as it requires white-box access to the LLM's internal parameters (e.g., vocabulary, gradients), which is inconsistent with our black-box attacker model. Instead, for the NQ and HotpotQA datasets, malicious content is generated by appending incorrect answers to the questions. For the domain-specific MediNote and Finance Alpaca datasets, we manually craft and store fixed malicious content for each domain within the RAG system; (b) \textbf{PoisonedRAG Attack}, which retrieves target answers to dynamically instantiate context-specific deceptive content.

\textbf{Defender Settings}: We compare our method against three other paradigms under identical environmental constraints: 
a) \textbf{SS-RAG}, a baseline configuration with a single malicious source and no defenses; 
b) \textbf{MS-RAG}, a multi-source extension with adversarial participation below 50\%; and 
c) \textbf{RobustRAG} and d) \textbf{FilterRAG}. Their respective mechanisms are applied individually to all documents from each source. For FilterRAG, the keyword density threshold is set to 0.2, following its ablation study.
All configurations preserve identical adversarial conditions and parameters throughout evaluation cycles.
For training SCR model, we adopt common learning settings. We employ AdamW optimizer with a weight decay of 1e-5. A CyclicLR scheduler adjusts the learning rate between a base of 1e-5 and a maximum of 1e-3. As previously described, our total loss function is composed of a classification loss, a topology-aware loss, and a gate regularization term. The topology loss weight $\lambda$ is adjusted via cosine annealing from 1 to 0. The gate regularization term imposes an $\ell_1$-norm penalty on gate parameters to enforce sparsity, with its coefficient $\alpha$ set to 0.01. For a detailed implementation, please refer to the \textbf{code in our GitHub repository}\footnote{\url{https://github.com/Aquarids/scr-pub}}.

\subsection{Results and Analysis}

Table~\ref{tab:defense_performance_time} presents the ASR across different datasets and attacks, which constitutes our primary analytical focus. The presented metrics originate from 50 independent executions, each comprising 50 question-answer rounds, with final success rates calculated after excluding minimal instances of LLM response parsing failures. 
The experimental results, summarized in Table~\ref{tab:defense_performance_time}, present three key findings. First and foremost, the SCR framework consistently achieves the low Attack Success Rates across all tested configurations. On the MediNote dataset with Deepseek-V3, for example, SCR reduced the ASR against PoisonedRAG to 5.24\%, while the ASRs for MS-RAG and RobustRAG were 78.80\% and 80.12\%. Further analysis reveals that while architectural choices provide a foundational defense—as evidenced by the ASR against PoisonedRAG on MediNote dropping from 97.00\% (SS-RAG) to 78.80\% (MS-RAG) due to content dilution—this protection is insufficient on its own. Moreover, the ATA attack, a static method that relies on substituting a few keywords within otherwise normal text, is less potent against baseline systems than the dynamic PoisonedRAG (e.g., 78.68\% vs. 97.00\% ASR on SS-RAG). However, its subtle nature makes it more effective against SCR in several scenarios (e.g., 5.72\% ASR for ATA vs. 5.24\% for PoisonedRAG on MediNote with SCR). This suggests that the minor, static modifications of ATA may generate fewer detectable anomalies for defenses.

\begin{table}[h]
\centering
\caption{SCR Accuracy (\%) on MediNote (Deepseek-V3) with and without commonsense mechanism.}
\label{tab:commonsense_ablation}
\small
\begin{tabular}{l|ccc}
\hline
\textbf{Metric} & \textbf{SAA} & \textbf{PoisonedRAG} & \textbf{ATA} \\
\hline
\textbf{TPR (w/)} & 95.00 & 93.14 & 97.02 \\
\textbf{TPR (w/o)} & 88.40 & 89.96 & 88.99 \\
\hline
\textbf{TNR (w/)} & 94.33 & 99.75 & 93.62 \\
\textbf{TNR (w/o)} & 86.50 & 93.33 & 80.45 \\
\hline
\textbf{FPR (w/)} & 5.67 & 0.25 & 6.38 \\
\textbf{FPR (w/o)} & 13.50 & 6.67 & 19.55 \\
\hline
\textbf{FNR (w/)} & 5.00 & 6.86 & 2.98 \\
\textbf{FNR (w/o)} & 11.60 & 10.03 & 11.01 \\
\hline
\end{tabular}
\end{table}

To ensure the effectiveness of the commonsense mechanism, we conduct experiments on the MediNote dataset using the Deepseek-V3 model, with results reported in Table~\ref{tab:commonsense_ablation}. The data shows that incorporating commonsense reasoning enhances SCR's performance across all metrics. For instance, under the SAA attack, TPR improves from 88.40\% to 95.00\% and TNR from 86.50\% to 94.33\% with the mechanism enabled, alongside substantial reductions in FPR and FNR.
Even without the commonsense mechanism, the SCR framework exhibits considerable robustness, with TPR consistently above 88\% and TNR generally above 80\%. Meanwhile, a closer inspection reveals that the ATA attack presents a challenge due to its stealthiness. Without the commonsense mechanism, ATA results in the lowest TNR at 80.45\%, lower than PoisonedRAG's TNR of 93.33\%, indicating its more covert nature.

\begin{table}[h]
\centering
\caption{Average Source Reliability (0-1) with SecureCollaRAG}
\label{tab:source_reliability_single}
\begin{tabular}{l|cc|cc}
\hline
\textbf{Attack}
& \multicolumn{2}{c|}{MediNote} 
& \multicolumn{2}{c}{Finance Alpaca} \\
\cline{2-5}
& Benign & Malicious & Benign & Malicious \\
\hline
SAA & 0.983 & 0.009 & 0.998 & 0.012 \\
PoisonedRAG   & 0.885 & 0.018 & 0.852 & 0.016 \\
ATA           & 0.818 & 0.287 & 0.793 & 0.246 \\
\hline
\end{tabular}
\end{table}
Table~\ref{tab:source_reliability_single} demonstrates SecureCollaRAG's capability to effectively distinguish between attack types through accumulated interaction rounds. The framework shows highly effective separation between benign and malicious sources for SAA attacks (e.g., 0.983 vs. 0.009 on MediNote). While PoisonedRAG attacks yield low malicious scores (0.018), ATA attacks achieve the highest malicious reliability scores (0.287 on MediNote and 0.246 on Finance Alpaca), confirming their enhanced stealth characteristics. These progressive divergences in reliability scores confirm the system's adaptive discrimination capacity across varied adversarial patterns.
Based on the above conclusions, our SCR framework demonstrates its effectiveness in balancing security and practicality.

\section{Conclusion}
Our research focuses on mitigating the critical vulnerability of RAG systems to knowledge corruption attacks. We introduce a novel defense framework, \textbf{SecureCollaRAG}, which constructs a multi-source document graph and leverages dynamic graph neural networks to compute document credibility, enabling robust multi-source knowledge validation. Additionally, we propose the Adaptive Tampering Attack, a sophisticated adversarial strategy utilizing prompt engineering to achieve better stealth. Evaluations demonstrate SCR's robustness against attackers under non-IID data distributions, establishing a new paradigm for secure multi-agent collaboration.

\bibliographystyle{IEEEtran}
\bibliography{secure_colla_rag}

\appendix

\subsection{Detailed Feature Descriptions}
\label{app:feature_details}

SecureCollaRAG's node features incorporate comparative metrics enhanced by a logarithmic normalization scheme to prevent domination by features with large value ranges and to better capture relative differences:
\begin{equation}
f(x, \tilde{x}) = \ln\left(1 + \frac{x}{\tilde{x}}\right)
\label{eq:logarithmic_norm}
\end{equation}
where $x$ denotes the local measurement from a single document and $\tilde{x}$ represents the global median value for that feature across all documents in the current batch. The 12 discriminative signals are as follows:

\textbf{Document Content Features}:
\begin{itemize}
    \item \textbf{Content length ratio}: Applies logarithmic scaling \eqref{eq:logarithmic_norm} on the document's content length to assess its verbosity against the typical length in the specific domain.
    \item \textbf{Keywords score ratio}: Computes a TF-IDF-based summation over relevant n-grams. The score computation employs a linguistic processing pipeline: input text undergoes initial tokenization using specialized stopwords from \texttt{nltk}, followed by systematic generation of 1-3 gram combinations constrained to 2-35 character phrases with a minimum of 6 cumulative letters per expression. This multi-stage filtering produces meaningful n-grams while automatically excluding generic terms. The final score is normalized using \eqref{eq:logarithmic_norm}.
    \item \textbf{Keywords count ratio}: Applies logarithmic scaling \eqref{eq:logarithmic_norm} to the count of extracted keywords relative to the global median.
    \item \textbf{Category distribution ratio}: Measures the frequency of a document's assigned category within the current submissions. This is achieved by computing linguistic similarity across categorical descriptions (e.g., \textit{"Unknown"}: \textnormal{"Unclassified or unspecified medical documents"}, \textit{"Cardiology"}: \textnormal{"Medical specialty concerning heart disorders"}) to detect anomalous or rare category assignments.
    \item \textbf{Sentiment polarization}: A four-dimensional VADER analysis captures emotional intensity through: (1) positive valence quantifying explicit optimism, (2) neutral ratio indicating emotional ambivalence, (3) negative magnitude measuring critical sentiment, and (4) a compound score reflecting the overall emotional orientation.
\end{itemize}

\textbf{Graph Structural Features}:
\begin{itemize}
    \item \textbf{Edge Density Metric}: Defined as $\frac{\text{deg}(v_i)}{|\mathcal{V}|}$, this feature measures a node's connectivity density scaled by the graph size. Higher values may indicate an attempt by the node to disproportionately influence the verification consensus.
    \item \textbf{Degree Ranking}: Transforms the raw node degree into a logarithmic rank using \eqref{eq:logarithmic_norm}. This transformation demotes the influence of high-degree hub nodes while amplifying subtle structural differences that are critical for detecting collusion patterns.
    \item \textbf{Semantic Cohesion}: Integrates global and local semantic patterns through a weighted logarithmic combination. The global semantic similarity is scaled by a factor $\beta$ and combined with the local neighborhood similarity (computed as the average of top-$k$ most similar nodes) via $(1-\beta)$. This dual-scale mechanism uses logarithmic compression to balance system-wide semantic trends with localized contextual relationships, mitigating distortion from extreme similarity values.
\end{itemize}

\textbf{Cross Interaction Features}:
\begin{itemize}
    \item \textbf{Lexical-Semantic Relevance}: Fuses term frequency statistics (lexical information) with contextual relationships (semantic information) to identify inconsistencies that often appear in adversarial content, such as plausible but contextually irrelevant statements.
\end{itemize}

\subsection{Detailed BFT Analysis and Proofs}
\label{app:bft_proofs}

As visualized for the PoisonedRAG attack scenario in Fig.~\ref{fig:visual_1}, we observe a distinct clustering pattern. The t-SNE projection of document feature vectors (left panel) reveals a clear separation between malicious documents (red circles) and benign ones (green squares), which are submitted by different participants. This phenomenon stems from the fact that documents generated with a shared malicious objective exhibit high similarity. As explained by the autoregressive nature of LLMs \cite{factoscope}, similar input prefixes constrain the decoding process toward analogous token probability distributions, leading to outputs that are closely clustered in the feature space.

Visual analysis of PoisonedRAG (Fig.~\ref{fig:visual_1}) reveals two notable patterns: (1) A distinct separation boundary exists between malicious and benign documents within query groups, confirming cluster compactness hypotheses; (2) Malicious documents demonstrate higher spatial concentration compared to their benign counterparts.
Building on the separation patterns observed, we formalize SecureCollaRAG's robustness under the constraint $|\mathcal{S}_m|/|\mathcal{S}| \leq \rho < 0.5$. Let $\mathcal{D}_m^{(t)}$ and $\mathcal{D}_b^{(t)}$ denote malicious and benign document sets at round~$t$. The instantaneous risk combines false acceptance and rejection probabilities:

\begin{equation}
\begin{aligned}
R_t := & \underbrace{\Pr\!\left[d \in \mathcal{D}_m^{(t)} \wedge \mathrm{C}(d) \geq \theta_{c}\right]}_{\text{false acceptance}} \\
& + \underbrace{\Pr\!\left[d \in \mathcal{D}_b^{(t)} \wedge \mathrm{C}(d) < \theta_{c}\right]}_{\text{false rejection}} .
\end{aligned}
\end{equation}

According to the t-SNE visualization, we assume cluster separation with margin $\Delta>0$ in the GNN's latent space~$\mathcal{Z}$:
\begin{equation}
\min_{\substack{d_m \in \mathcal{D}_m^{(t)} \\ d_b \in \mathcal{D}_b^{(t)}}} \| \phi(d_m) - \phi(d_b) \|_{\mathcal{Z}} \ge \Delta
\end{equation}
with the GNN approximating the Bayes-optimal classifier $C^\star$ within error $\eta$:
\begin{equation}
\sup_{d \in \mathcal{D}^{(t)}} |C(d) - C^\star(d)| \le \eta < \tfrac12 .
\end{equation}

\subsection{Lemma~1 (Cluster Separability)}\label{lemma:separable}
Let $\mathbf{H}^{(t)}=\{\mathbf{h}_i^{(t)}\}_{i=1}^{|\mathbb{D}^{(t)}|}$ be the set of node representations produced by the dynamic graph neural network at round~$t$.  
Assume the fraction of malicious sources satisfies $|\mathcal{S}_m|/|\mathcal{S}|\le \rho<0.5$ and the GNN has reached training equilibrium under the composite loss in~\eqref{eq:topo_loss}.  
Then there exists a decision function $f(\mathbf{h})$
and constants $\gamma>0$ (inter-cluster margin) and $\varepsilon\in(0,1)$ (tolerable error) such that

\begin{equation}
\Pr\!\left[
\begin{array}{l}
d\in\mathcal{D}_m^{(t)},\;f(\mathbf{h}_d^{(t)})\le -\gamma \\
\text{or}\\
d\in\mathcal{D}_b^{(t)},\;f(\mathbf{h}_d^{(t)})\ge +\gamma
\end{array}
\right]\;\;\ge 1-\varepsilon,
\end{equation}

i.e.\ malicious and benign documents are separable with margin~$\gamma$ up to probability~$1-\varepsilon$.

\paragraph{Proof.} 
The topology-aware loss~\eqref{eq:topo_loss} consists of two complementary terms:  
(i) the variance regularizer $\mathbb{V}\mathrm{ar}[\cos(\mathbf{h}_u,\mathbf{h}_v)]$, which contracts intra-cluster dispersion, and  
(ii) the cluster-contrastive component, which enlarges inter-cluster distances by maximizing the softmax-scaled similarity gap between representations and their respective centroids $\{\boldsymbol{\mu}_k\}$.  
At convergence, these terms enforce the following empirical property with high probability ($1-\delta$, $\delta\!<\!\varepsilon$):  

\begin{equation}
\underbrace{
    \max_{d_1,d_2\in\mathcal{D}_b^{(t)}} 
    \|\mathbf{h}_{d_1}^{(t)}-\mathbf{h}_{d_2}^{(t)}\|_2
}_{\text{benign radius}}
\le \sigma_b,
\end{equation}

\begin{equation}
\underbrace{
    \max_{d_1,d_2\in\mathcal{D}_m^{(t)}}
    \|\mathbf{h}_{d_1}^{(t)}-\mathbf{h}_{d_2}^{(t)}\|_2
}_{\text{malicious radius}}
\le \sigma_m,
\end{equation}

\begin{equation}
\begin{aligned}
&\underbrace{
    \min_{d_b\in\mathcal{D}_b^{(t)},\,d_m\in\mathcal{D}_m^{(t)}}
    \|\mathbf{h}_{d_b}^{(t)}-\mathbf{h}_{d_m}^{(t)}\|_2
}_{\text{inter-cluster gap}}
\\
&\ge \sigma_b+\sigma_m+\gamma,
\end{aligned}
\end{equation}

for some radii $\sigma_b,\sigma_m>0$ and margin $\gamma>0$.  
(1) follows from the bounded variance term, while (2) is guaranteed by the contrastive push-apart force combined with the assumption $\rho<0.5$, which prevents malicious points from dominating the representation space.

Because the benign and malicious clusters are enclosed in disjoint $\ell_2$-balls of radii $\sigma_b$ and $\sigma_m$ that are separated by at least $\gamma$, the classical Fisher discrimination result ensures the existence of a hyperplane $\{\mathbf{h}\mid f(\mathbf{h})=0\}$ with margin~$\gamma/2$ that cleanly separates the two closed balls.  
The probability that an unseen sample violates the margin is bounded by $\delta$ due to concentration---treating DGNN as a deterministic Lipschitz mapping, standard Rademacher generalization bounds yield an additional failure term $<\varepsilon-\delta$.  

Combining the two failure probabilities gives $\Pr[\text{error}]\le\varepsilon$, concluding that the DGNN representations are separable within error~$\varepsilon$. \qed

\subsection{Lemma~2 (Calibration Error Bound)}\label{lemma:calib}

Let $\sigma(x)=\bigl(1+\mathrm{e}^{-x}\bigr)^{-1}$ be the logistic sigmoid and recall the decision function $f(\mathbf{h})=\mathbf{w}^{\!\top}\mathbf{h}+b$ from Lemma~\ref{lemma:separable}.  
Define the \emph{credibility score} of a document $d$ at round~$t$ as
\begin{equation}
\mathrm{C}\!\left(d^{(t)}\right)=\sigma\!\bigl(f(\mathbf{h}_d^{(t)})\bigr).
\end{equation}
The \emph{expected calibration error} (ECE) is
\begin{equation}
\operatorname{ECE}_t=
\mathbb{E}_{d \sim \mathbb{D}^{(t)}}
\Bigl|
\,\mathrm{C}\!\left(d^{(t)}\right)
-\mathbf{1}\!\{d\in\mathcal{D}_b^{(t)}\}
\Bigr|.
\end{equation}
Under the conditions of Lemma~\ref{lemma:separable},
\begin{equation}
\operatorname{ECE}_t\le
\varepsilon+(1-\varepsilon)\,\mathrm{e}^{-\gamma}.
\end{equation}

\paragraph{Proof.}
Lemma~\ref{lemma:separable} guarantees that, with probability at least $1-\varepsilon$,
\begin{equation}
\begin{aligned}
&d\in\mathcal{D}_b^{(t)} \;\Longrightarrow\; f(\mathbf{h}_d^{(t)})\ge\gamma, \\
&d\in\mathcal{D}_m^{(t)} \;\Longrightarrow\; f(\mathbf{h}_d^{(t)})\le-\gamma.
\end{aligned}
\end{equation}
Consider a correctly separated benign document.  Monotonicity of $\sigma$ gives
\begin{equation}
\begin{aligned}
\mathrm{C}\!\left(d^{(t)}\right) 
&= \sigma\!\bigl(f(\mathbf{h}_d^{(t)})\bigr) \\
&\ge \sigma(\gamma) \\
&= \frac{1}{1+\mathrm{e}^{-\gamma}} \\
&= 1 - \frac{\mathrm{e}^{-\gamma}}{1+\mathrm{e}^{-\gamma}} \\
&\ge 1-\mathrm{e}^{-\gamma},
\end{aligned}
\end{equation}
hence
\begin{equation}
\Bigl|\,\mathrm{C}\!\left(d^{(t)}\right)-1\Bigr|\le\mathrm{e}^{-\gamma}.
\end{equation}
For a correctly separated malicious document we analogously have
\begin{equation}
\Bigl|\,\mathrm{C}\!\left(d^{(t)}\right)-0\Bigr|\le\mathrm{e}^{-\gamma},
\quad\text{since}\quad
\sigma(-\gamma)\le\mathrm{e}^{-\gamma}.
\end{equation}

Therefore, on the high-probability event of correct separation,
the absolute calibration error does not exceed $\mathrm{e}^{-\gamma}$.
On the complementary event of probability $\varepsilon$ no guarantee exists, so the worst-case error is bounded trivially by~$1$.  Aggregating the two cases,
\begin{equation}
\operatorname{ECE}_t
\le
(1-\varepsilon)\,\mathrm{e}^{-\gamma}+\varepsilon\cdot1
=\varepsilon+(1-\varepsilon)\,\mathrm{e}^{-\gamma}.
\end{equation}
Because $\varepsilon\in(0,1)$ and $\gamma>0$, the bound lies strictly below~$1$ and decreases exponentially in~$\gamma$, confirming that larger inter-cluster margins yield better probabilistic calibration. \qed

\subsection{Lemma~3 (Reliability Convergence)}\label{lemma:reliability}

For a fixed source $s_j\!\in\!\mathcal{S}$ define the \emph{instantaneous average credibility}  
\begin{equation}
\label{eq:avg_cred}
\bar{C}_j^{(t)}
=\frac{1}{\lvert\mathcal{D}_j^{(t)}\rvert}
\sum_{d\in\mathcal{D}_j^{(t)}}\mathrm{C}\!\left(d^{(t)}\right),
\end{equation}
and let $C_j^{\star}\!=\!\lim_{t\to\infty}\bar{C}_j^{(t)}\in[0,1]$ exist.  
Then the exponentially-smoothed reliability score  
$\varrho^{(t)}(s_j)$ generated by  
\begin{equation}
\label{eq:ewma_update}
\varrho^{(t+1)}(s_j)=\beta\,\varrho^{(t)}(s_j)
+(1-\beta)\,\bar{C}_j^{(t)},
\quad 0<\beta<1,
\end{equation}
converges to the same limit $C_j^{\star}$ with the error bound
\begin{equation}
\label{eq:conv_bound}
\begin{aligned}
\bigl|\varrho^{(t)}(s_j)-C_j^{\star}\bigr|
\;\le\;&\;
\beta^{t}\bigl|\varrho^{(0)}(s_j)-C_j^{\star}\bigr| \\
&+
(1-\beta)\sum_{u=0}^{t-1}\beta^{u}
\bigl|\bar{C}_j^{(t-1-u)}-C_j^{\star}\bigr|.
\end{aligned}
\end{equation}
In particular, if $\bigl|\bar{C}_j^{(t)}-C_j^{\star}\bigr|\!\xrightarrow[t\to\infty]{}\!0$, then $\varrho^{(t)}(s_j)\!\xrightarrow[t\to\infty]{}\!C_j^{\star}$.

\paragraph{Proof.}
Define the error term $e_t=\varrho^{(t)}(s_j)-C_j^{\star}$.  
Subtracting $C_j^{\star}$ from both sides of~\eqref{eq:ewma_update} yields
\begin{equation}
e_{t+1}
=\beta\,e_t
+(1-\beta)\bigl(\bar{C}_j^{(t)}-C_j^{\star}\bigr).
\end{equation}
Iteratively unrolling the recurrence gives
\begin{equation}
e_{t}
=\beta^{t}e_0
+(1-\beta)\sum_{u=0}^{t-1}\beta^{u}
\bigl(\bar{C}_j^{(t-1-u)}-C_j^{\star}\bigr),
\end{equation}
from which the absolute bound in~\eqref{eq:conv_bound} follows by the triangle inequality.  
Since $0<\beta<1$ the pre-factor $\beta^{t}$ decays exponentially;  
convergence of $\bar{C}_j^{(t)}$ implies the summand  
$\bigl|\bar{C}_j^{(t-1-u)}-C_j^{\star}\bigr|$ vanishes as $t\!\to\!\infty$, forcing the entire sum to disappear.  
Consequently $\lim_{t\to\infty}e_t=0$, i.e.\ $\varrho^{(t)}(s_j)\to C_j^{\star}$. \qed

\subsection{Proof of Theorem~\ref{theorem:risk}.}

Partition the sample space into the \emph{separable event} $\mathcal{E}$, where the margin guarantees of Lemma~\ref{lemma:separable} hold, and its complement $\bar{\mathcal{E}}$.  
By definition  
\begin{equation}
\Pr(\bar{\mathcal{E}})\;\le\;\varepsilon,\quad
\Pr(\mathcal{E})=1-\varepsilon.
\end{equation}

\textit{Risk on} $\bar{\mathcal{E}}$.  
No structural guarantee is available; applying the trivial bound yields  
\begin{equation}
R_t\mid \bar{\mathcal{E}}\;\le\;1.
\end{equation}

\textit{Risk on} $\mathcal{E}$.  
Within this event every benign document satisfies $f(\mathbf{h})\!\ge\!\gamma$ and every malicious one $f(\mathbf{h})\!\le\!-\gamma$.  
Lemma~\ref{lemma:calib} therefore implies  
\begin{equation}
\mathrm{C}(d_b)\;\ge\;1-\mathrm{e}^{-\gamma},
\quad
\mathrm{C}(d_m)\;\le\;\mathrm{e}^{-\gamma}.
\end{equation}
Because the threshold $\theta_{c}$ lies strictly between these two values, both false-acceptance and false-rejection probabilities vanish inside $\mathcal{E}$, so  
\begin{equation}
R_t\mid\mathcal{E}\;=\;0.
\end{equation}
Combining the two cases yields
\begin{equation}
\label{eq:Rt_decompose}
\begin{aligned}
R_t
&= \Pr(\bar{\mathcal{E}})\,R_t\!\mid\!\bar{\mathcal{E}}
  + \Pr(\mathcal{E})\,R_t\!\mid\!\mathcal{E} \\
&\le \Pr(\bar{\mathcal{E}})\cdot 1 + \Pr(\mathcal{E})\cdot 0 \\
&= \varepsilon .
\end{aligned}
\end{equation}
Substituting the calibration bound $R_t\!\mid\!\bar{\mathcal{E}}\le \mathrm{e}^{-\gamma}$ in place of the trivial~$1$ strengthens~\eqref{eq:Rt_decompose} to the form stated in~\eqref{eq:risk_bound}.

Finally, Lemma~\ref{lemma:reliability} ensures that sources with persistently low average credibility receive exponentially diminishing weights, causing the prevalence of malicious documents in $\mathbb{D}^{(t)}$ to contract over time.  
Consequently the contribution of the failure event $\bar{\mathcal{E}}$ asymptotically vanishes, yielding the limit in~\eqref{eq:risk_limit}.  \qed

\subsection{Visual Analysis of Adaptive Tampering Attack}
\label{appendix:ata_analysis}

The Adaptive Tampering Attack achieves fundamental improvements through three key divergence mechanisms. Unlike PoisonedRAG's predictable outputs, ATA maintains attack effectiveness while generating diverse malicious documents. First, it operates on naturally distributed benign documents rather than creating artificial clusters of malicious content, enabling better camouflage. Second, a randomized tactic selector produces different attack patterns even for identical concepts through varied output generation paths. Third, ATA separates attack preparation from execution by pre-generating poisoned content, avoiding the detectable timing patterns seen in PoisonedRAG's real-time attacks.

\begin{figure*}[h]
   \centering
   \includegraphics[width=.7\linewidth]{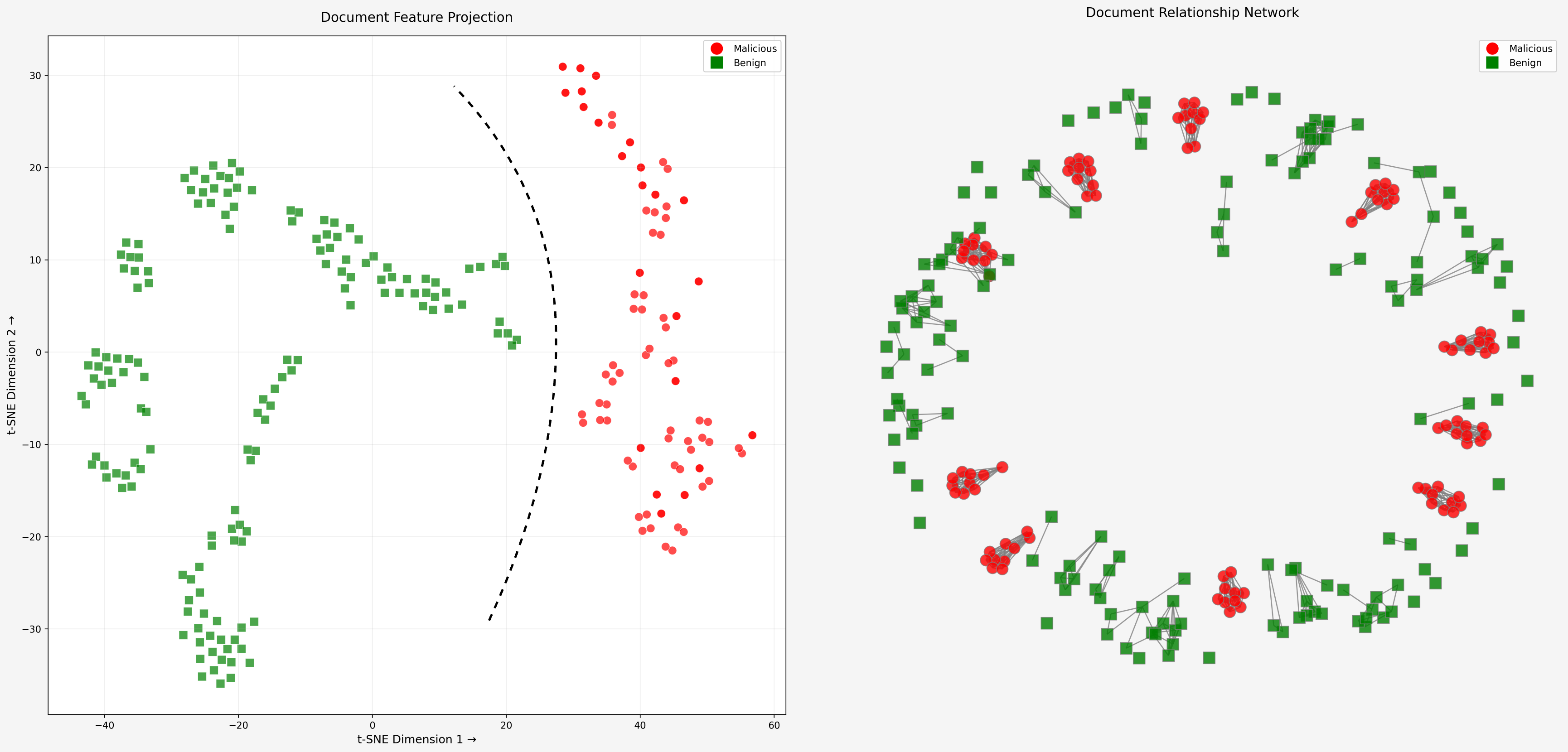}
   \caption{Visual analysis of PoisonedRAG.}
   \label{fig:visual_1}
\end{figure*}

\begin{figure*}[h]
   \centering
   \includegraphics[width=.7\linewidth]{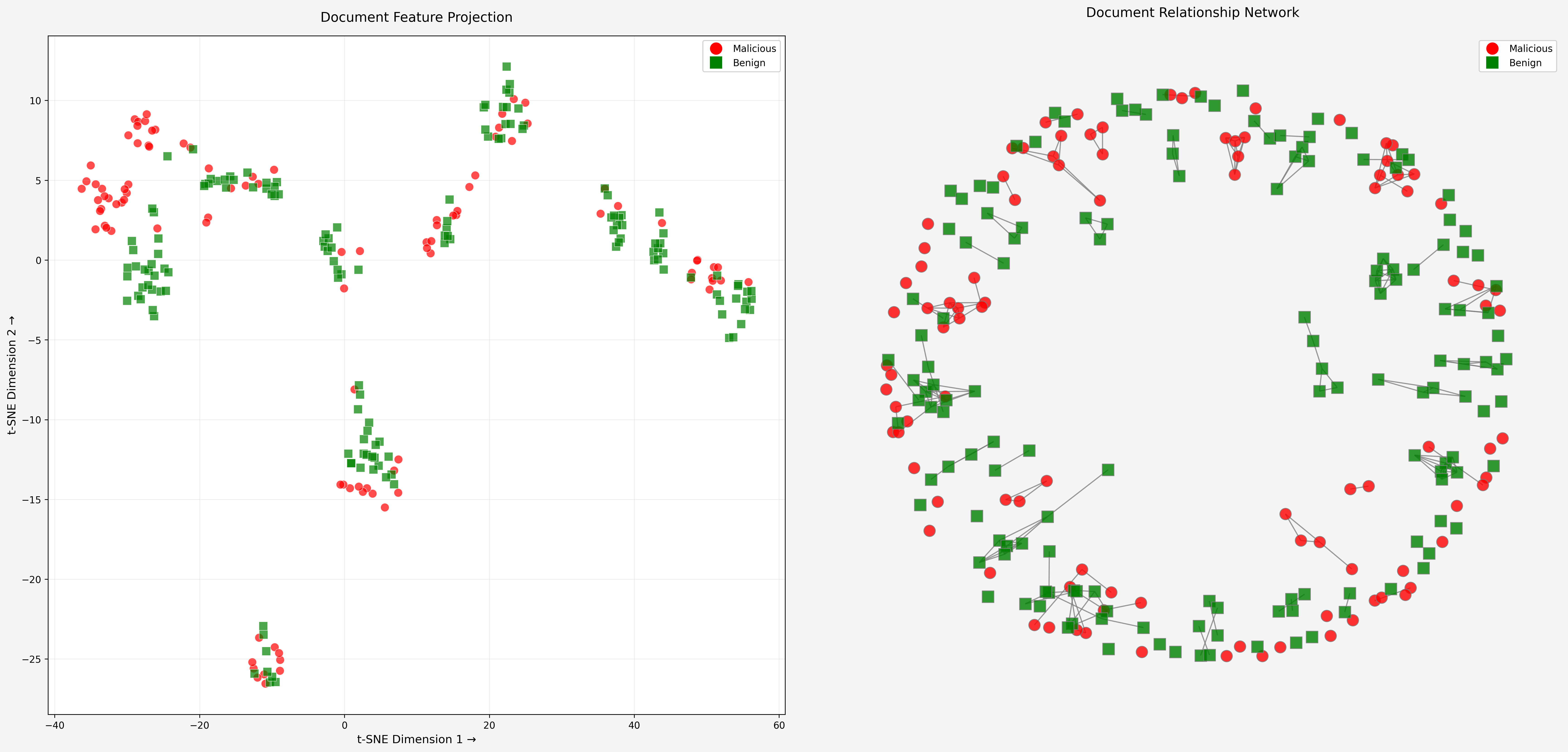}
   \caption{Visual analysis of the Adaptive Tampering Attack.}
   \label{fig:visual_2}
\end{figure*}

Although this pre-generation of malicious content may not always precisely target specific questions for poisoning as effectively as PoisonedRAG, its inherent stealthiness is enhanced, making it much harder for defense mechanisms to detect. To validate this claim, we provide comparative t-SNE visualizations in Fig.~\ref{fig:visual_1} and Fig.~\ref{fig:visual_2}. As shown in the document feature projections, PoisonedRAG exhibits a clear linear separation boundary between malicious (red) and benign (green) documents, making them easily distinguishable. In contrast, ATA-generated malicious documents are substantially more dispersed and intermingled with benign samples, demonstrating significantly reduced detectability. These innovations allow malicious documents to blend with legitimate content through following aspects: 1) leveraging authentic document distributions, 2) introducing controlled randomness in attack patterns, and 3) erasing signatures in attack execution. The combined effect breaks traditional detection methods that rely on content similarity or timing analysis, effectively erasing the clear boundaries between adversarial and benign outputs.

\subsection{Graph Neural Network and Graph Attention Networks}

Graph Neural Networks (GNNs) formalize relational reasoning through distributed state transition dynamics \cite{gnn}. Let $G=(V,E)$ define a graph with vertex set $V$ and edge set $E$, where each node $v_i \in V$ maintains a state vector $\mathbf{h}_i^{(t)} \in \mathbb{R}^d$ at iteration $t$. The information propagation mechanism follows:

\begin{equation}
\mathbf{h}_i^{(t+1)} = \sigma\left(\mathbf{W}_{\text{self}}\mathbf{h}_i^{(t)} + \sum_{j \in \mathcal{N}(i)}\mathbf{W}_{\text{neigh}}\mathbf{h}_j^{(t)}\right),
\end{equation}

where $\mathcal{N}(i)$ denotes the neighborhood of node $v_i$, $\mathbf{W}_{\text{self}}$ and $\mathbf{W}_{\text{neigh}}$ are learnable parameters, and $\sigma$ represents a nonlinear activation function.

Graph Attention Networks (GATs) enhance this paradigm through adaptive neighborhood weighting. The attention coefficient $\alpha_{ij}$ between nodes $v_i$ and $v_j$ is computed as:

\begin{equation}
\alpha_{ij} = \frac{\exp\left(\text{ReLU}\left(\mathbf{a}^\top[\mathbf{W}\mathbf{h}_i \| \mathbf{W}\mathbf{h}_j]\right)\right)}{\sum_{k \in \mathcal{N}(i) \cup \{i\}}\exp\left(\text{ReLU}\left(\mathbf{a}^\top[\mathbf{W}\mathbf{h}_i \| \mathbf{W}\mathbf{h}_k]\right)\right)},
\end{equation}

where $\mathbf{a}$ and $\mathbf{W}$ are learnable parameters, and $\|$ denotes vector concatenation. The final node representation aggregates neighborhood features through these attention weights:

\begin{equation}
\mathbf{h}_i' = \sigma\left(\sum_{j \in \mathcal{N}(i) \cup \{i\}}\alpha_{ij}\mathbf{W}\mathbf{h}_j\right).
\end{equation}

\enlargethispage{3\baselineskip}
This architecture enables differentiable importance weighting of neighbor nodes without requiring prior structural knowledge \cite{gat}.

\end{document}